\documentclass[twocolumn,english,prl,superscriptaddress,showpacs,amssymb]{revtex4}
\usepackage[T1]{fontenc}
\usepackage{amsmath}
\usepackage{amssymb}
\usepackage{graphicx}

\makeatletter
\@ifundefined{textcolor}{}
{%
 \definecolor{BLACK}{gray}{0}
 \definecolor{WHITE}{gray}{1}
 \definecolor{RED}{rgb}{1,0,0}
 \definecolor{GREEN}{rgb}{0,1,0}
 \definecolor{BLUE}{rgb}{0,0,1}
 \definecolor{CYAN}{cmyk}{1,0,0,0}
 \definecolor{MAGENTA}{cmyk}{0,1,0,0}
 \definecolor{YELLOW}{cmyk}{0,0,1,0}
 }

\usepackage{bm}\@ifundefined{definecolor}{\usepackage{color}}{}

\usepackage[T1]{fontenc}
\usepackage[latin9]{inputenc}
\usepackage{textcomp}
\usepackage{color}

\usepackage{babel}

\makeatother

\usepackage{babel}
\begin{document}

\title{Generating Entangled Microwave Radiation Over Two Transmission Lines}

\author{E. Flurin}

\affiliation{Laboratoire Pierre Aigrain, Ecole Normale Sup\'erieure, CNRS (UMR 8551),
Universit\'e P. et M. Curie, Universit\'e D. Diderot 24, rue Lhomond,
75231 Paris Cedex 05, France }

\author{N. Roch}

\affiliation{Laboratoire Pierre Aigrain, Ecole Normale Sup\'erieure, CNRS (UMR 8551),
Universit\'e P. et M. Curie, Universit\'e D. Diderot 24, rue Lhomond,
75231 Paris Cedex 05, France }

\author{F. Mallet}

\affiliation{Laboratoire Pierre Aigrain, Ecole Normale Sup\'erieure, CNRS (UMR 8551),
Universit\'e P. et M. Curie, Universit\'e D. Diderot 24, rue Lhomond,
75231 Paris Cedex 05, France }

\author{M. H. Devoret}

\affiliation{Coll\`ege de France, 11 Place Marcelin Berthelot, F-75231 Paris Cedex
05, France }

\affiliation{Laboratoire Pierre Aigrain, Ecole Normale Sup\'erieure, CNRS (UMR 8551),
Universit\'e P. et M. Curie, Universit\'e D. Diderot 24, rue Lhomond,
75231 Paris Cedex 05, France }

\affiliation{Department of Applied Physics, Yale University, PO Box 208284, New
Haven, CT 06520-8284 }

\author{B. Huard}

\email[corresponding author: ]{benjamin.huard@ens.fr}

\selectlanguage{english}%

\affiliation{Laboratoire Pierre Aigrain, Ecole Normale Sup\'erieure, CNRS (UMR 8551),
Universit\'e P. et M. Curie, Universit\'e D. Diderot 24, rue Lhomond,
75231 Paris Cedex 05, France }

\date{\today}

\begin{abstract} \textbf{Using a superconducting circuit, the Josephson mixer, we demonstrate the first experimental realization of spatially separated two-mode squeezed states of microwave light. Driven by a pump tone, a first Josephson mixer generates, out of quantum vacuum, a pair of entangled fields at different frequencies on separate transmission lines. A second mixer, driven by a $\pi$-phase shifted copy of the first pump tone, recombines and disentangles the two fields. The resulting output noise level is measured to be lower than for vacuum state at the input of the second mixer, an unambiguous proof of entanglement. Moreover, the output noise level provides a direct, quantitative measure of entanglement, leading here to the demonstration of 6~Mebit.s$^{-1}$ (Mega entangled bits per second) generated by the first mixer.}\end{abstract}

\maketitle

Pairs of entangled electromagnetic fields propagating on physically separated channels constitute an essential resource in quantum information processing, communication and measurements \cite{Braunstein:2005p7586,Weedbrook:2011p7584}. They can be realized by squeezing a vacuum state shared by two spatially separated modes. This entanglement
is revealed in the cross-correlations between well chosen quadratures of the two fields which fall below the level of quantum vacuum noise.
Given the considerable development of microwave quantum optics, these Einstein-Podolsky-Rosen (EPR) states, or spatially separated two-mode squeezed vacuum states, have become highly desirable at such frequencies. At optical frequencies, EPR states are usually prepared by parametric down-conversion of a pump
tone using a $\chi^{(2)}$ nonlinear medium \cite{SLUSHER:1985p7579,Wenger:2005p7780}. 
At microwave frequencies, only single-mode squeezing and two-mode squeezing 
between sidebands of a single transmission line have been demonstrated
so far, using degenerate Josephson parametric amplifiers \cite{yurke_observation_1988,CastellanosBeltran:2009p2186,Mallet:2011p7670,Eichler:2011p7784,Wilson}.
Recently, a dissipationless, nondegenerate, three-wave mixer for
microwave signals based on Josephson junctions was developed \cite{Bergeal:2010p6952,Bergeal:2012p6951,Roch:2012p7781} (see Figs.~1,2).
Strong correlations between the spontaneously emitted radiations from
two ports have been observed in the parametric down-conversion mode
\cite{Bergeal:2010p7251}, but the experiment did not prove directly
the presence of entanglement in the separated output fields. 

Here,
we describe an interference experiment demonstrating that nondegenerate
Josephson mixers can entangle and disentangle usable EPR states of
microwave light (Fig. 1). A first mixer, called the {}``entangler\textquotedblright{},
is driven by a pump tone while its two input ports are terminated
by cold loads ensuring that only vacuum quantum noise enters the device.
The two entangled output ports feed the input ports of a second
mixer called the {}``analyzer\textquotedblright{}. The role of the
analyzer is to recombine and disentangle the two microwave fields before sending
them to a standard microwave amplification and detection chain.
As the phase difference between both pumps varies, the noise at the output of the analyzer exhibits interference fringes which pass under the level of amplified vacuum.
Remarkably, the measurement of the noise at the output of the analyzer directly quantifies entanglement between its two
input fields without resorting to two homodyne detection channels and the analysis of their correlations.

\begin{figure*}
\includegraphics[scale=0.57]{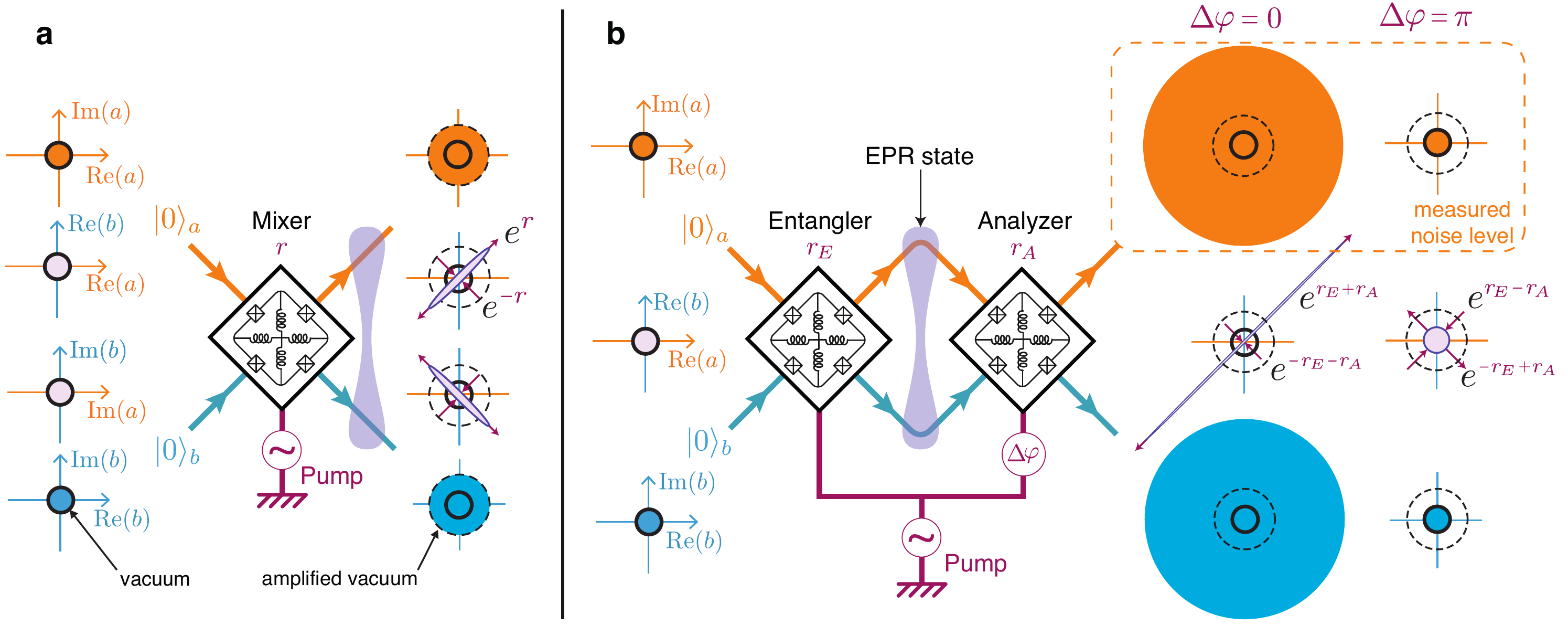} \caption{\textbf{Principle of the experiment. a.} When pumped with a microwave tone at frequency $f_{P}=f_{a}+f_{b}$, a Josephson mixer (black diamond) transforms incoming quantum vacuum noise (left) on modes $a$ and $b$ into an EPR state (right). The field states are represented by their standard deviation contours in the single mode phase space of $a$ (top), $b$ (bottom) and in the bipartite phase space (middle) spanned by $(\mathrm{Re}(a),\mathrm{Re}(b))$ and $(\mathrm{Im}(a),\mathrm{Im}(b))$ where $\mathrm{Re}(a)=(a+a^{\dagger})/2$ and $\mathrm{Im}(a)=(a-a^{\dagger})/2i$ are the in-phase and out-of-phase quadratures of mode $a$. In each plot, a solid black circle sets the scale of vacuum noise and a dashed circle sets the scale of amplified vacuum noise. Quantum entanglement in the output fields is observed in the bipartite phase space where cross-correlations go beyond quantum uncertainty by a squeezing factor $e^{-r}$.  \textbf{b.} In order to demonstrate entanglement at the ouput of this mixer named {}``entangler'', a second, identical mixer named {}``analyzer'' is placed in series and pumped by the same tone with phase difference $\Delta\varphi$. Entanglement at the input of the analyzer is revealed by measuring, at the analyzer output (dashed frame), a lower noise level on mode $a$ or $b$  than for amplified vacuum (dashed circle) for a given phase $\Delta\varphi$. The smallest (largest) output noise level occurs for opposite (equal) squeezing factor at $\Delta\varphi=\pi$ (0), and can ideally go as low as the vacuum noise level.}
\end{figure*}

The Josephson mixer \cite{Bergeal:2010p6952,Bergeal:2012p6951,Roch:2012p7781} is a superconducting
circuit parametrically coupling two superconducting resonators (Fig.
2) at distinct frequencies $f_{a}$ and $f_{b}$ via a pump
at their sum frequency $f_{P}=f_{a}+f_{b}$. Each resonator has only
one access port, but input and output signals are spatially separated
by cryogenic microwave circulators (Fig. 2 and \cite{Supinfo}) so that the entangler
output can be exclusively sent to the analyzer input. Each mixer performs
a reversible transform of the wavefunction of the field via the unitary
two-mode squeeze operator $S=\exp\left(re^{i\varphi_{P}}a^{\dagger}b^{\dagger}-re^{-i\varphi_{P}}ab\right)$
where $re^{i\varphi_{P}}$ is the complex squeezing parameter, and $a$ and $b$ are the field operators of the two modes \cite{Grynbergbook}.
The input and output canonical field operators are related by the
scattering relations 
\begin{equation}
\begin{array}{rll}
a_{out} & =S^{\dagger}a_{in}S= & \cosh(r)a_{in}+e^{i\varphi_{P}}\sinh(r)b_{in}^{\dagger}\\
b_{out}^{\dagger} & =S^{\dagger}b_{in}^{\dagger}S= & \cosh(r)b_{in}^{\dagger}+e^{-i\varphi_{P}}\sinh(r)a_{in}
\end{array}
\label{scattering}
\end{equation}
where $\varphi_{P}$ is the phase of the pump and $G=\cosh^{2}r=(P_{th}+P)^{2}/(P_{th}-P)^{2}$
is the power direct gain which increases with pump power $P$ below the parametric self-oscillation threshold $P_{th}$. With the
pump on, the vacuum state at the input is converted into a two-mode
squeezed vacuum state $\left|Sq\right\rangle =S\left|0\right\rangle _{a}\left|0\right\rangle _{b}=\cosh(r)^{-1}\sum\tanh(r)^n\left|n\right\rangle _{a}\left|n\right\rangle _{b}$.
Note that this entangled state can be understood as the superposition of twin
photons with different frequencies and propagating on spatially
separated transmission lines. Non-local two-mode squeezing directly appears
in the combinations of output fields
\begin{equation}
a_{out}\pm e^{i\varphi_{P}}b_{out}^{\dagger}=e^{\pm r}\left(a_{in}\pm e^{i\varphi_{P}}b_{in}^{\dagger}\right)
\end{equation}
which, for $\varphi_{P}=0$, implies cross-correlations between $\mathrm{Re}(a)$
and $\mathrm{Re}(b)$ on one hand and $\mathrm{Im}(a)$ and $-\mathrm{Im}(b)$ on the other
hand, beating the Heisenberg limit of vacuum quantum noise, as shown
in Fig. 1. In optics, these correlations have been observed by double
balanced homodyne detection techniques in several systems \cite{Reid:2009p7776}.
The present experiment describes the first demonstration at microwave
frequencies of these quantum correlations between signals on spatially separated
transmission lines. The Josephson mixer here serves two functions. First, the entangler produces EPR states of microwave light at incommensurate frequencies and on spatially
nondegenerate modes with squeezing parameter $r_{E}$. Second, the analyzer recombines input fields as shown in Eq.~(\ref{scattering}), with squeezing parameter $r_{A}$ and relative pump phase $\Delta\varphi$, in order to reverse the transformation and disentangle the field state (Fig. 1).

The output noise of the entangler can be measured on each mode independently by turning off the analyzer $(r_A=0)$. The noise power spectrum measured by a spectrum analyzer is proportional
to the symmetrized variance of the field operator \cite{Clerk:2010p7260}
\begin{equation}
(\Delta a_{out,E})^{2}=\frac{\left\langle \left\{ a_{out,E},a_{out,E}^{\dagger}\right\} \right\rangle}{2} -\left|\left\langle a_{out,E}\right\rangle \right|^{2}=\frac{\cosh2r_{E}}{2}. \label{vacinput}
\end{equation}
The variance of this "amplified vacuum" is always larger than that of the vacuum state, for
which $(\Delta a)^{2}=1/2$ (Fig. 1). 
Discarding the information from the other mode, each output field is in a thermal state \cite{Drummond:2004p7794}. Yet, since the combined
two-field state $\left|Sq\right\rangle $ is a pure state with no entropy,
it is possible, ideally, to reverse the squeezing with a second mixer and re-obtain
a vacuum state on each port. The analyzer can perform this
inversion if operated with opposite squeezing parameter $r_{A}=-r_{E}$.
In practice, unavoidable losses between the two mixers prevent
the exact recovery of the vacuum.

\begin{figure}
\includegraphics[scale=0.6]{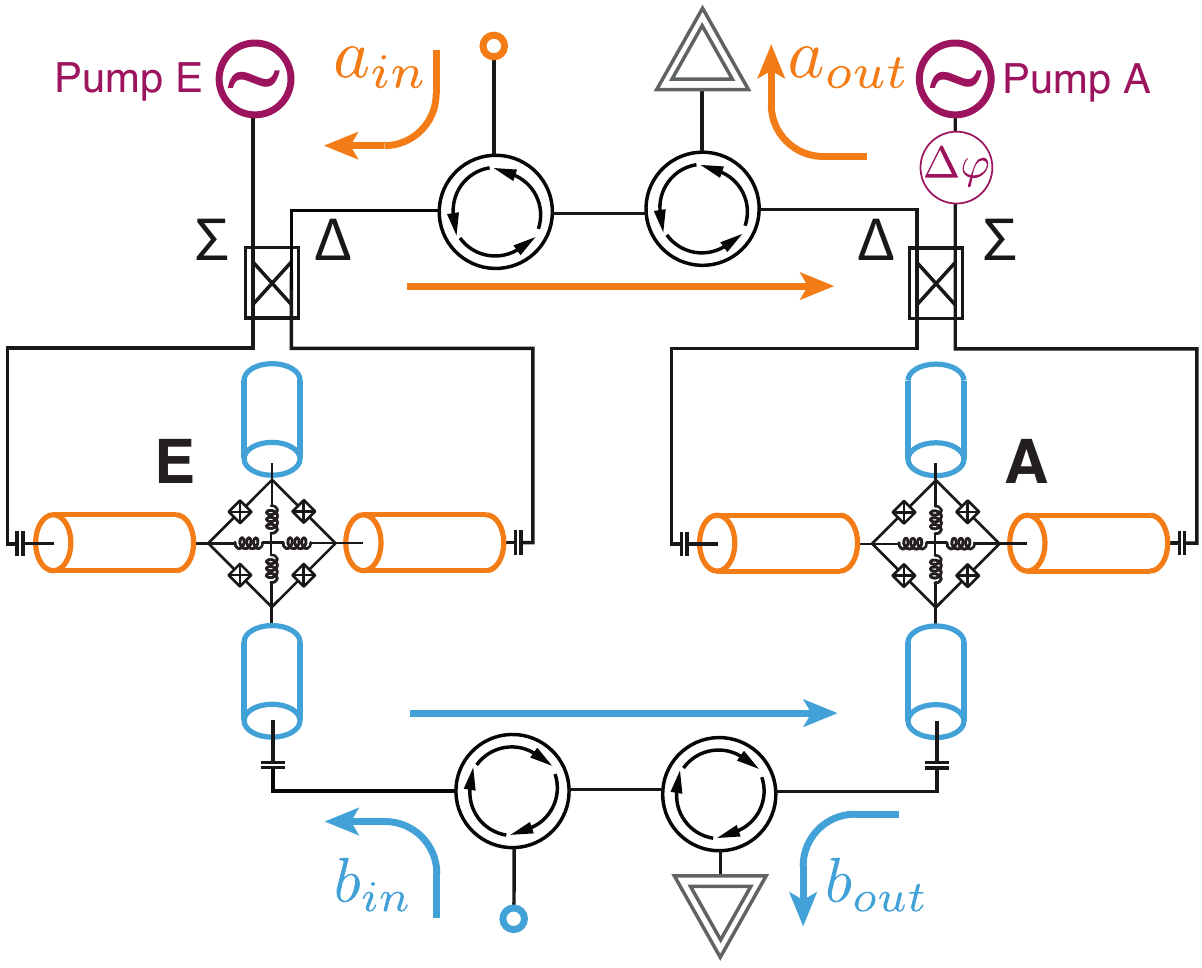}\caption{Schematics of the experimental setup. Each Josephson mixer consists of a ring of Josephson junctions coupling
two $\lambda/2$ superconducting resonators addressed via a 180\textdegree{}
hybrid coupler or a single ended port. Both mixers are designed with the same geometry as in Ref. \cite{Roch:2012p7781}, and their resonance frequencies are matched at $f_{a}=5.578\,$GHz and $f_{b}=8.812\,$GHz using two independent flux bias. The pump frequencies are set to $f_P=14.390~\mathrm{GHz}$. Microwave circulators
separate the input and output of the entangler and analyzer. Input ports are represented as open circles, and 
at each output port, the double triangle symbolizes the low noise amplifying measurement setup with total gain $G_\mathrm{LNA}$.
}
\end{figure}

Losses are modeled as field splitters coupling uncorrelated cold thermal
baths to each mode (Fig. 3) so that \begin{equation*}a_{in,A}=\sqrt{\bar{\alpha}}a_{out,E}+\sqrt{\alpha}a_{th}\textrm{ and }b_{in,A}=\sqrt{\bar{\beta}}b_{out,E}+\sqrt{\beta}b_{th}\end{equation*}
where $a_{th}$ and $b_{th}$ describe bosonic modes of thermal
baths at frequencies $f_{a}$ and $f_{b}$, and $\bar{\alpha}=1-\alpha$, $\bar{\beta}=1-\beta$. Besides, microwaves photons
propagate for a finite amount of time $\tau_{a}$ and $\tau_{b}$
between the two mixers leading to a correction of the phase difference
entering the scattering terms $\Delta\varphi'=\Delta\varphi-2\pi f_{a}\tau_{a}-2\pi f_{b}\tau_{b}.$
The temporal extent of the twin photons exiting the entangler is given
by the inverse of the bandwidth $\Delta f=\Delta f_{0}/\cosh r_{E}$ \cite{Bergeal:2010p7251}.
In the experiment, the travel times $\tau_{a}$ and $\tau_{b}$ of
order 2 ns are much smaller than this temporal extent since $\Delta f_{0}=28\textrm{ MHz}$,
so that microwave photons do interfere even if their travel durations
may slightly differ between modes. It is then straightforward to calculate
the scattering coefficients of the full circuit. For instance, the $a$ output mode is given by 
\begin{eqnarray*}
a_{out,A}&=&t_{a\rightarrow a}a_{in,E}+t_{b\rightarrow a}b_{in,E}^{\dagger}\\&&+\sqrt{\alpha}\cosh r_{A}a_{th}+e^{i\Delta\varphi}\sqrt{\beta}\sinh r_{A}b_{th}^{\dagger}.
\end{eqnarray*}
where
\begin{equation}
\begin{array}{rcl}
t_{a\rightarrow a} & = & \sqrt{\bar{\alpha}}\cosh r_{E}\cosh r_{A}+e^{i\Delta\varphi}\sqrt{\bar{\beta}}\sinh r_{E}\sinh r_{A},\\
t_{b\rightarrow a} & = & \sqrt{\bar{\alpha}}\sinh r_{E}\cosh r_{A}+e^{i\Delta\varphi}\sqrt{\bar{\beta}}\cosh r_{E}\sinh r_{A}.\end{array}\label{eq:fullSparam}\end{equation}

These scattering coefficients were measured using a nonlinear four-port vector network analyzer as a function of the phase difference $\Delta\varphi$ for various values of the gains
$\cosh^{2}r_{E,A}$ ranging from 1 to 40, a subset of which is shown on Fig.
3. The special cases where one or both of the converters are not pumped
$(r=0)$ offer the opportunity to calibrate each converter gain independently.
The only fit parameter for this whole set of measurements
is the ratio between transmissions on both arms, found to be $\bar{\beta}/\bar{\alpha}=0.945$.
\begin{figure}
\includegraphics[scale=0.46]{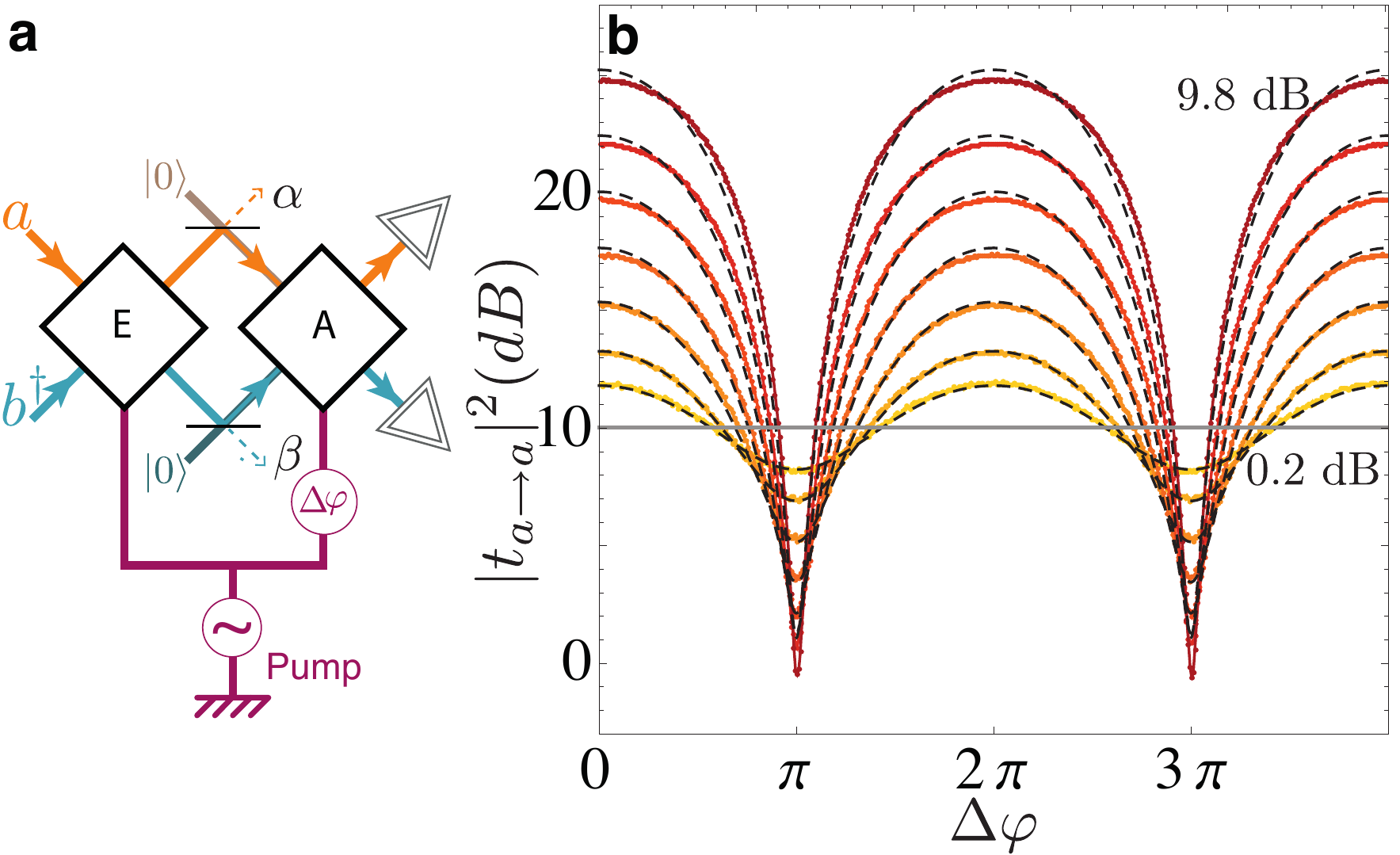}\caption{\textbf{a.} Protocol of the scattering coefficient measurements
by a vector network analyzer connected
between the $a,b$ input and the $a,b$ output ports. The setup is calibrated by
turning on and off each Josephson mixer separately. Losses are modeled as field splitters of transparency $\alpha^2$ and $\beta^2$ coupling a cold load to the signals.  \textbf{b.} Color traces: Transmission
measurements of $|t_{a\rightarrow a}|^{2}$ as a function of phase
difference $\Delta\varphi$ between both pump signals. The gain of
the analyzer is set to $G_{A}=\cosh^{2}r_{A}=10$ (solid gray line).
Each trace and color corresponds to a different gain for the entangler
$G_{E}=\cosh^{2}r_{E}=0.2,0.8,1.8,3.2,5,7.2,9.8$ dB. Dashed lines:
fits to the data using equation (\ref{eq:fullSparam}) and the single
fit parameter $\bar{\beta}/\bar{\alpha}=0.945$. Together with an independent, \emph{in situ} noise calibration this value leads to $\alpha=0.33\pm0.05$ and $\beta=0.36\pm0.05$ \cite{Supinfo}. }
\end{figure}
\begin{figure}
\includegraphics[scale=0.44]{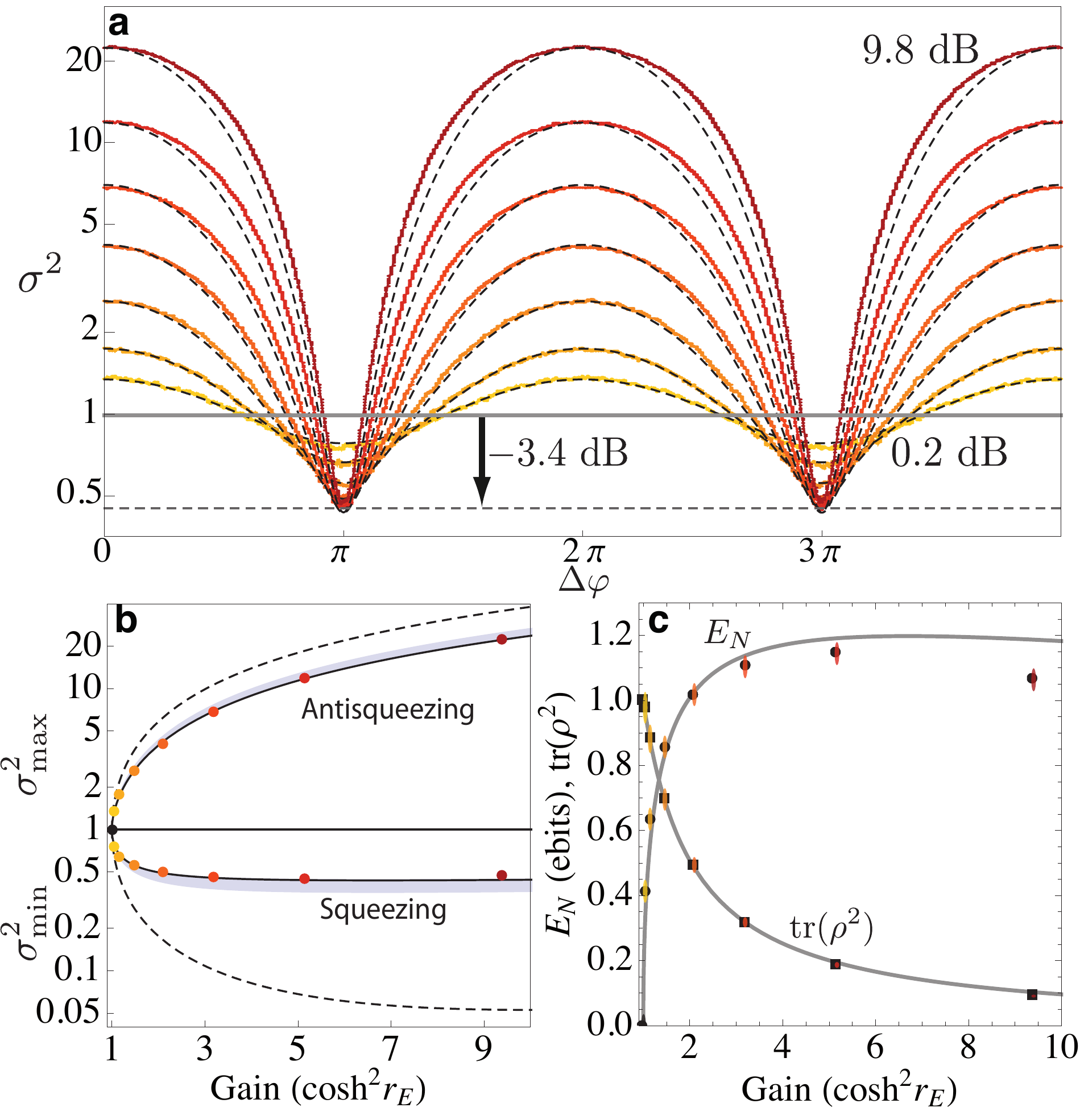}\caption{\textbf{a.} Color traces: variance of the
output mode $(\Delta a_{outA})^{2}$ referred to the case of vacuum input on the
analyzer (divided by $\cosh(2r_{A})/2)$ as a function of phase difference
$\Delta\varphi$, determined by measuring the spectral density of the noise at the analyzer a output when only quantum noise enters the entangler a,b inputs. An absolute calibration allows exact conversion between both quantities with an error of at most $\pm2.5\%$ \cite{Supinfo}. Each color corresponds
to the same gain of the entangler $G_E$ as in Fig~3 with a fixed gain on the analyzer $G_A=\cosh^{2}r_{A}=10$. The horizontal line at $\sigma^2=1$ represents the measured
noise for amplified vacuum at the output of the analyzer $(r_{E}=0)$.
For $\Delta\varphi$ close to $\pi$, the measured noise goes below this
level, an evidence of entanglement. Dashed lines: predicted
variance using Eq. (\ref{eq:cosphi}) extended to the unbalanced loss
case using $\alpha=0.37$ and $\beta=0.40$. \textbf{b.} Dots: Noise level measured at $\Delta\varphi=0$ (anti-squeezing)
and $\Delta\varphi=\pi$ (squeezing) as a function of gain $G_E$ for $G_A=10$. The size of the dots is larger than the error bar. Solid lines: prediction using
Eq. (\ref{eq:variance_series_cosphi=00003D00003Dpm1}), extended to
unbalanced losses as in panel 4a. Colored area: consistent values of the noise using the uncertainty in the calibration of the losses $\alpha$ and $\beta$ \cite{Supinfo}. Dashed lines: same prediction
but without losses, $\alpha=\beta=0$. \textbf{c.}  solid dots: logarithmic negativity measure of entanglement, with errors bars. solid squares: entanglement purity. Lines: theoretical predictions using the parameters of panel b.
}
\end{figure}

In Fig. 3, only mean values of the output field amplitudes are measured.
Yet, truly quantum features appear in their correlations. Consider
the case of a cold load setting the vacuum quantum state at the input
of the entangler, which is reached in our experiment at $45$~mK since $hf_{b}/k>hf_{a}/k=260\textrm{ mK}$ \cite{Supinfo}.
When the entangler is turned off ($r_E=0$),  the analyzer is fed by vacuum fluctuations and the output noise reads the amplified vacuum level $(\Delta a_{out,A})^{2}=\cosh(2r_{A})/2$ as in Eq.~(\ref{vacinput}). In general, the output noise $\sigma^2$, normalized to that reference level, on both output ports can be calculated from Eq.~(\ref{eq:fullSparam}) and oscillates with phase $\Delta\varphi$ as
\begin{equation}
\begin{array}{rcl}
\sigma^2(\Delta\varphi)&\equiv&2(\Delta a_{out,A})^{2}/\cosh (2r_A)\\&=&\bar{\alpha}\left(\cosh2r_{E}+\sinh2r_{E}\tanh2r_{A}\cos\Delta\varphi\right)+\alpha.\label{eq:cosphi}
\end{array}
\end{equation}
For simplicity, this expression is only given in the case of balanced losses $\alpha=\beta$, but the general case
can be treated without much difficulty. 
The maximal and minimal values of $\sigma^2$, corresponding to the extrema of cross-correlation between quadratures of the EPR state (ellipses in Fig. 1a), are obtained respectively for $\Delta\varphi=0$ and $\Delta\varphi=\pi$
\begin{equation}
\sigma^2_{\mathrm{max},\mathrm{min}}=(1-\alpha)\frac{\cosh(2r_{E}\pm2r_{A})}{\cosh(2r_A)}+\alpha.\label{eq:variance_series_cosphi=00003D00003Dpm1}
\end{equation}
The existence of a phase $\Delta\varphi$ for which
the output noise $\sigma^2$ goes below 1, which demonstrate correlations beyond quantum uncertainty, is a sufficient
evidence of entanglement \cite{Supinfo,Giedke:2003p7792,Horodecki:2009p7798}.

The normalized noise power $\sigma^{2}$ is obtained by measuring the spectral density $S_{a}$ (detailed in supplementary information \cite{Supinfo})
\begin{equation}
\sigma^2(\Delta\varphi)=\dfrac{2}{\cosh(2r_A)} \left( \frac{S_a(\Delta\varphi)-S_\mathrm{off}}{hf_a G_\mathrm{LNA}}+\frac{1}2\right),  \label{eqtousecal}
\end{equation}
where the noise background due to the following amplifiers $S_\mathrm{off}$ is small
enough to be precisely subtracted.
The spectral densities $S_{a}(f_{a})$ and $S_{b}(f_{b})$ of both
modes at the output of the analyzer were measured using a microwave
spectrum analyzer behind a cryogenic low-noise preamplifier on a 
$0.5\,$MHz bandwidth. This bandwidth was chosen to be smaller than that of
the mixers for all combinations of gains $G_{E,A}=\cosh^{2}r_{E,A}$ and phase
differences $\Delta\varphi$. Importantly, it was possible to calibrate the total gain of the measurement setup $G_\mathrm{LNA}$ so that the normalized noise power $\sigma^{2}$ is measured with at most $\pm2.5\%$ relative error \cite{Supinfo}. This calibration was performed by turning on a single mixer at a time and varying the temperature of a thermally decoupled input load on mode $a_{in,E}$. As a side result, the calibration provides the loss $\alpha=0.33\pm0.05$ between mixers on mode $a$  which, together with the ratio $\bar{\beta}/\bar{\alpha}=0.945$ from Fig.~3, leaves no unknown parameters in the experiment.

As can be seen on Fig. 4a, the
noise does pass below the threshold of amplified vacuum noise, hence
proving the existence of entanglement. Note that measurements on mode $b$ (not shown) gave similar results as expected. Note also that minimum noise $\sigma^2_\mathrm{min}$ occurs at $|r_E|<|r_A|$ and not at exactly opposite squeezing parameters. This deviation may be due to the beginning of a saturation of the analyzer mixer, corroborated by the slight deviations of the fits in Fig.~3. For each squeezing parameter
$r_{E}$, it is possible to extract the extrema of noise $\sigma^2_{\mathrm{min,max}}$
 from the curves of Fig. 4a. These extremal noise measurements
(Fig. 4b) are well described by Eq. (\ref{eq:variance_series_cosphi=00003D00003Dpm1})
generalized to unbalanced losses between modes with $\alpha=0.37$ and $\beta=0.40$, consistently with the calibration. It is even possible to account
for the whole dependence of the measured noise on phase difference
$\Delta\varphi$ by generalizing Eq. (\ref{eq:cosphi}) using the same parameters (Fig. 4b). The overall minimum for the measured noise is reached at $\cosh^2r_{E}\approx 5$ and reads 
$\sigma^2_\mathrm{min}=0.45\pm0.01$ with a corresponding maximum $\sigma^2_\mathrm{max}=11.9 \pm 0.1 $. 

It is remarkable that the amount of noise
at the output of a single port of the analyzer directly measures the
entanglement between the two input fields. In particular, the minimum
of output noise is linked to the logarithmic negativity $E_N=-\log_2(\sigma^2_\mathrm{min})=1.15\pm0.04$ and to the entropy of formation $E_F=0.69\pm0.03$ entangled bits (ebits, see \cite{Supinfo}) \cite{Supinfo,Vidal,Adesso,DiGuglielmo,Marian,Bennett:1996p7799}: the deeper the noise fringes, the larger the entanglement. The purity of the entangled state is also related to both extrema $\mathrm{tr}(\rho^2)=(\sigma^2_\mathrm{min}\sigma^2_\mathrm{max})^{-1}=0.186\pm0.09$.
These quantities of entanglement are within a factor of 2 from the state of the art in optics \cite{Laurat:2005p7801, DiGuglielmo, Takeno, Eberle}. Given the bandwidth of the mixers, the analyzer receives a usable rate of 6~Mebits.s$^{-1}$ from the entangler. 

In conclusion, we have demonstrated the production of EPR states of microwave radiation. Vacuum noise at the input of a first mixer is converted into two entangled
fields. A second mixer is used to recombine and disentangle the two fields. Using an absolute calibration, the minimal noise intensity at the output of the second mixer, when the phase difference $\Delta\varphi$
is varied, constitutes a direct measure of the entanglement between the twin
fields. Our measurements are limited by the finite losses between mixers
but still show that a rate of 6~Mebits.s$^{-1}$ travel between the entangler and the analyzer. This first implementation
of spatially separated two-mode squeezed states in the microwave domain opens novel experiments on quantum teleportation or superdense coding in the fields of nanomechanical resonators and superconducting qubits. Moreover, inserting a "circuit QED" readout cavity in one arm of the vacuum quantum noise interferometer described in this paper, one would achieve a maximally efficient measurement, for a given photon number, of the phase shift associated with a change of qubit state.

\begin{acknowledgments} We are grateful to N. Treps, T. Kontos and J.-P. Poizat for fruitful discussions, P. Morfin for technical assistance and P. Pari for giving us RuO$_2$ thermometers. Nanofabrication has been made
within the consortium Salle Blanche Paris Centre. This work was supported
by the EMERGENCES program Contract of Ville de Paris and by the ANR
contract ULAMSIG.\end{acknowledgments}

\bibliographystyle{apsrev}

\end{document}